\documentstyle[graphics,psfig]{caosp}
\begin{document}
\pubyear{1998}
\volume{27}
\firstpage{431}
\htitle{Modelling of magnetic fields in CP stars}
\hauthor{S.~Bagnulo}
\title{Broadband linear polarization and modelling of magnetic fields
       in CP stars}
\author{S.~Bagnulo}
\institute{Institut f\"{u}r Astronomie, T\"{u}rkenschanzstrasse 17, A-1180
           Wien, Austria}
%\date{}
\maketitle
\begin{abstract}
The observations in broadband linear polarization (BBLP) represent a
powerful diagnostic tool for detecting the magnetic
morphology of CP stars. In several cases, BBLP observations are not
consistent with the prediction of the oblique rotator model with a
simple dipolar field. Discrepancies between theoretical predictions
and observations may partially be ascribed to the effect of
desaturation due to the blending of spectral lines, which has not been
taken into account so far in the models. However, there are strong
evidences that in many cases the simple dipolar model is not
sufficient to describe the magnetic configuration of CP stars.

\end{abstract}

\section{Diagnostic techniques for detecting magnetic fields in CP
         stars}\label{Sect_Techniques}
The detection of magnetic fields in CP stars is based on the analysis
of the Zeeman effect in the Stokes parameter profiles of spectral
lines. Roughly speaking, the line intensity profile is sensitive to
the mean magnetic field modulus, whereas the spectrum of the radiation
observed in left and right circular polarization gives information on
the mean longitudinal magnetic field. The linear polarization profile
of a spectral line is sensitive to the transverse component of the
magnetic field. Magnetic fields can be detected with several
techniques, which have different constraints and provide different 
kinds of information.

From the relative wavelength shift of the line components split by
the magnetic field, one can derive the 
\textit{mean magnetic field modulus} $\langle\vert B \vert\rangle$,
that is, the average over the visible stellar disk of the modulus of
the magnetic field. So far, $\langle \vert B \vert \rangle$ has been
measured in 42 CP stars. About a dozen of them are well monitored
throughout the entire stellar rotation cycle (Mathys et al.\ 1997).

The wavelength shift of the spectral lines between right and left
circular polarization is proportional to the average over the stellar
disk of the component of the magnetic field along the line of sight, 
$\langle B_z \rangle$. This quantity is called 
\textit{mean longitudinal magnetic field}.
An alternative method to derive the same quantity is based on the
anaylsis of the wings of the H$_\beta$ line (Borra \& Landstreet 1980).
Most of our knowledge of magnetic fields in CP stars relies on the
observations of $\langle B_z \rangle$. Observations were performed
e.g., by Babcock (1958), Borra \& Landstreet (1980), Mathys
(1991). About 150 stars have been monitored so far.

A quantitative analysis of the cross-over effect allows one to derive
the average over the stellar disk of the component of the magnetic
field along the line of sight times the distance between the point on
the stellar surface and the plane defined by the line of sight and the
stellar rotation axis (Mathys 1995a). This quantity, named
\textit{mean asymmetry of the longitudinal magnetic field},
$\langle dB_z \rangle$, has been detected in 44 stars. About a dozen of
them were well monitored throughout the rotation cycle.

The second order moment of the profile $I$ about the line
centre is proportional to the sum of the squares of the two previous
quantities, $\langle B^2 + B^2_z \rangle$ (Mathys 1995b). 
This quantity, shortly called \textit{mean quadratic magnetic field},
has been detected in about 40 stars, 11 of them 
well monitored in the rotation cycle.

The transfer of radiation in the stellar atmosphere is responsible for
a differential saturation of the $\sigma$ and $\pi$ components of
the Zeeman multiplet, which gives rise to a phenomenon of
\textit{broadband linear polarization}. This phenomenon
has been known for a long time to be observable on sunspots (Leroy
1962), but only very recently this kind of observations have been
systematically performed on CP stars (Leroy 1996). The observed BBLP
is roughly proportional to some bilinear forms built up with the
transverse components of the magnetic field, namely, 
$\langle B^2_x - B^2_y \rangle$ and $\langle B_xB_y \rangle$,
where $xyz$ represents an orthogonal reference system with the $z$
axis oriented as the line of sight, and $x$ towards the north celestial
pole (Landolfi et al.\ 1993). BBLP has been detected in 55 CP stars, 16
of them are well monitored throughout the rotation cycle.
%%%%%%%%%%%%%%%%%%%%%%%%%%%%%%%%%%%%%%%%%%%%%%%%%%%%%%%%%%%%%%%%%%%

\section{The magnetic observable quantities and the oblique rotator
         model}\label{Sect_Observable}
Let us consider a star with a dipolar field. The oblique rotator model
(ORM) can be characterized by the angle between line of sight and
rotational axis, $i$ ($0^\circ \le i \le 180^\circ$), and the angle
between rotational axis and magnetic axis, $\beta$ ($0^\circ \le \beta
\le 180^\circ$). The observed quantities related to the magnetic
field may be calculated as a function of the stellar phase. 
The dotted lines in Fig.~1 show the various observable
quantities as predicted for some magnetic configurations. Instead of
plotting $Q$ and $U$ as a function of phase, it is more interesting to
plot the expected values of BBLP in  the $Q-U$ plane. The patterns
so obtained are called ``BBLP diagrams''.  

%%%%%%%%%%%%%%%%%%%%%%%%%%%%%%%%%%%%%%%%%%%%%%%%%%%%%%%%%%%%%%%%%%%
\begin{figure}[t]
\label{Fig_Observable}
%\scalebox{0.77}{
%\psfig{figure=T3f1.ps,width=11.8cm,clip=}
%\includegraphics*[2.5cm,8.2cm][18.cm,19.5cm]{Bagnulo_1.ps}}
\centerline{
\psfig{file=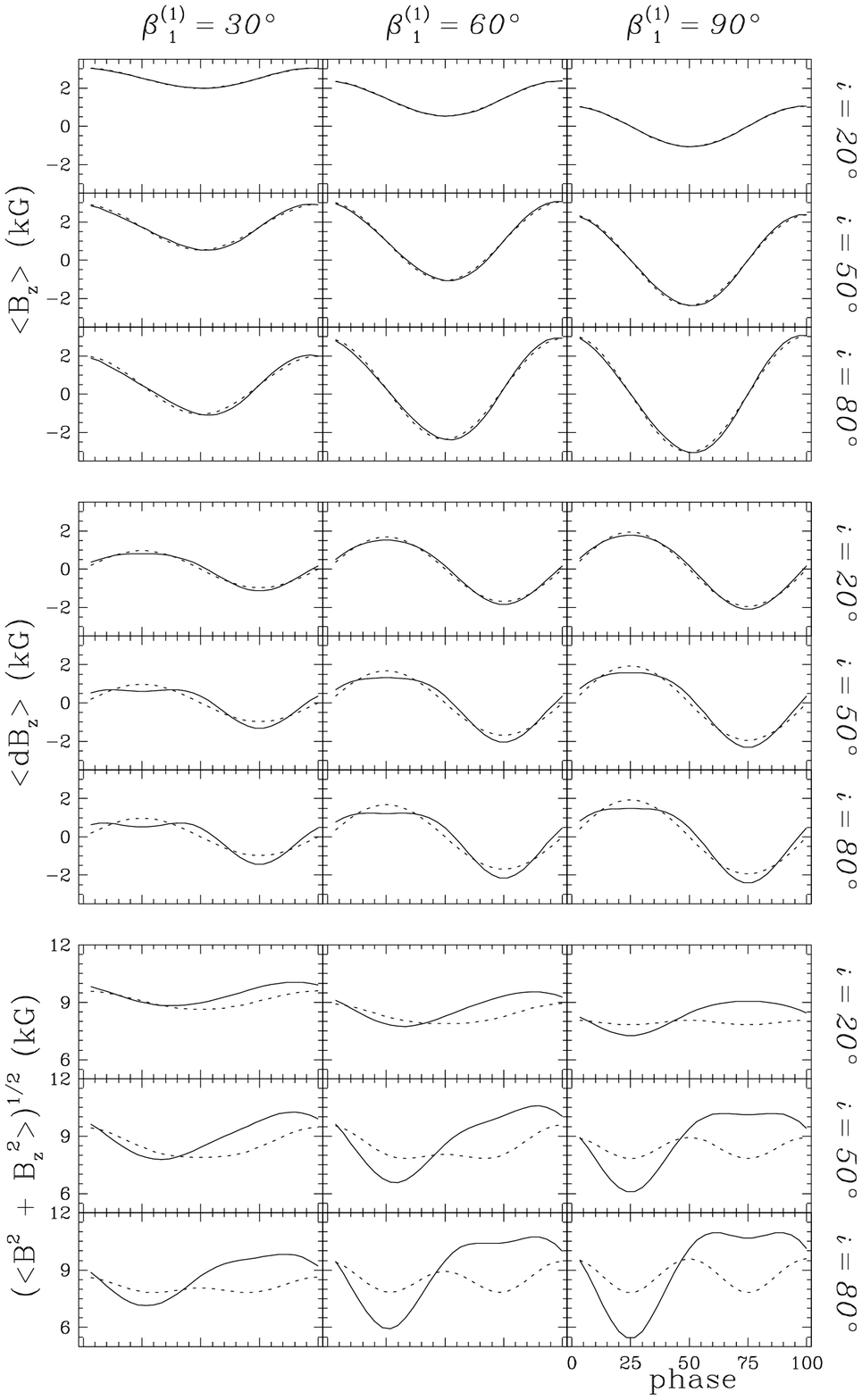,height=9.cm,clip=}
\hfill
\psfig{file=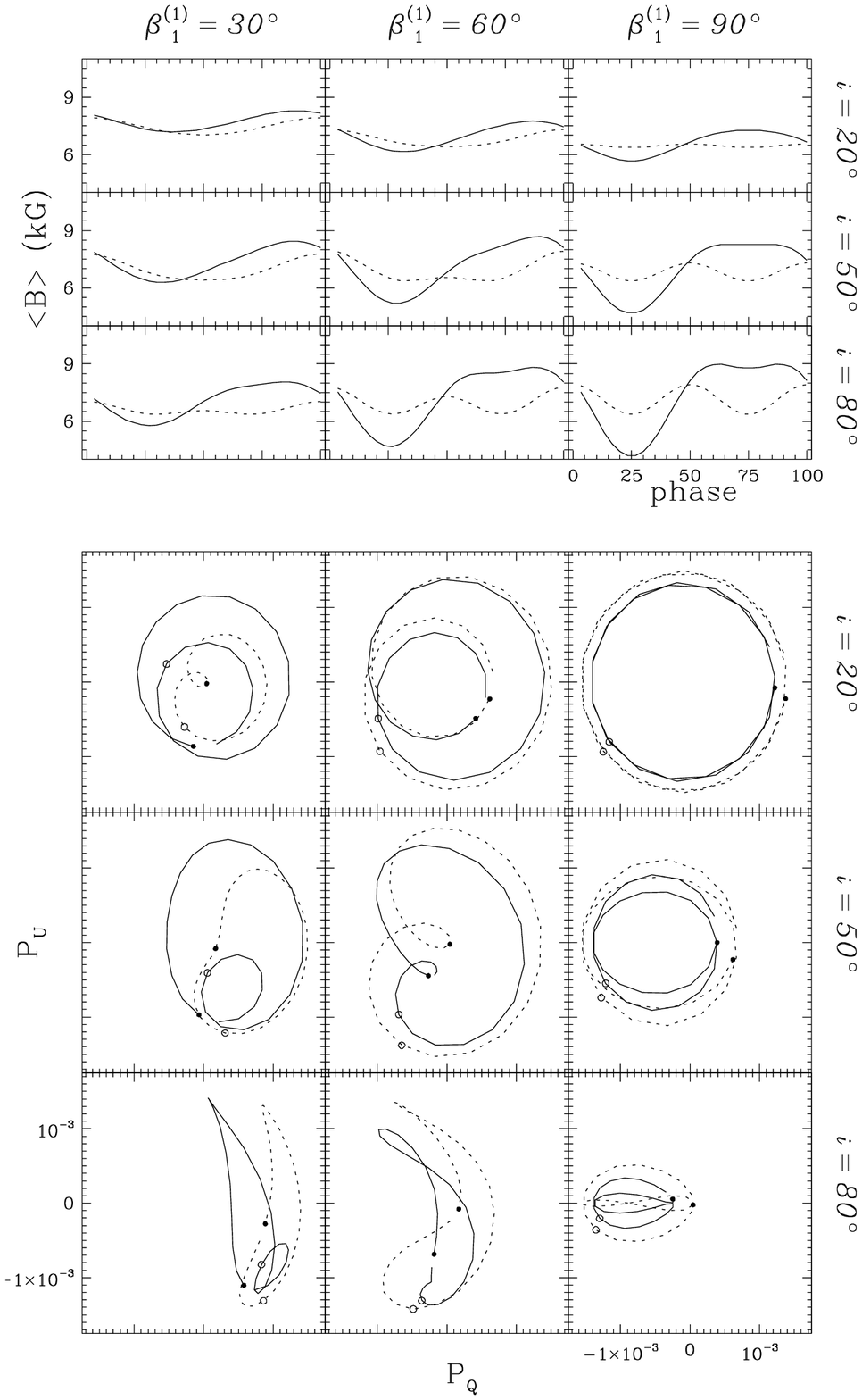,height=9.cm,clip=}
}
\caption{\small Dotted lines: the various observable quantities
calculated for the magnetic configurations corresponding to the ORM
with a dipolar field with a strength of 10\,kG, and with different
values of $i$ and $\beta$. Solid lines: the observable
quantities are calculated for the same magnetic configuration as for
the dotted line, but with the superposition of a quadrupole
perpendicular to the stellar rotational axis.
($\langle dB_z\rangle$ is calculated with Eq.~(69) of Bagnulo et al.\
(1996) with its sign changed in order to be consistent with the definition
of Mathys (1995a.)} 
\end{figure}
%%%%%%%%%%%%%%%%%%%%%%%%%%%%%%%%%%%%%%%%%%%%%%%%%%%%%%%%%%%%%%%%%%%

\section{Observations and modelling}\label{Sect_Observations}
Among 16 CP stars, in only one case
(HD~24712) the observations of BBLP are consistent with the
theoretical predictions (Bagnulo et al.\ 1995, Leroy 1996).
There are two strong assumptions that we have done in order
to derive the theoretical BBLP diagrams.  

So far, it has been assumed that the entire stellar spectrum
might be characterised by a single 
``typical'' line, rather than a more realistic model atmosphere
(Landolfi et al.\ 1993). Indeed, in many cases, the BBLP
does not depend strongly on the Zeeman patterns of spectral lines.
However, there is an effect of depolarization due to the 
blending of several spectral lines, which might affect the shape of the
BBLP diagrams in the presence of strong magnetic
fields. Figure~2 shows the BBLP for a Zeeman triplet
formed in a Milne-Eddington atmosphere (left) compared to the BBLP
calculated for a stellar atmosphere of 10\,000\,K. 
%%%%%%%%%%%%%%%%%%%%%%%%%%%%%%%%%%%%%%%%%%%%%%%%%%%%%%%%%%%%%%%%%%%
\begin{figure}[t]
\label{Fig_BBLP}
\begin{center}
\scalebox{0.6}{
\includegraphics*[2.5cm,8.2cm][18.cm,15.5cm]{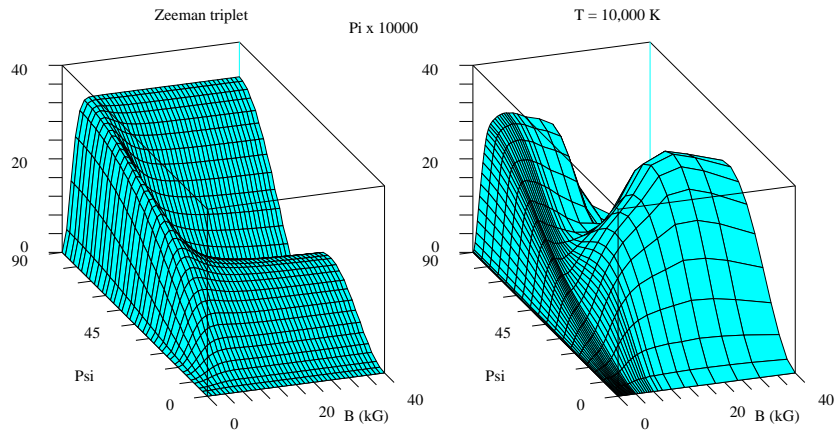}}
\end{center}
\caption{Left: the BBLP (in units per $10^{-4}$) calculated for a
normal Zeeman triplet formed in a Milne-Eddington atmosphere, as a
function of the modulus of the magnetic field (expressed in kG), and
the angle between the line of sight and the direction of the magnetic
field. Right: the same for a stellar model atmosphere (see Stift 1998)}
\end{figure}
%%%%%%%%%%%%%%%%%%%%%%%%%%%%%%%%%%%%%%%%%%%%%%%%%%%%%%%%%%%%%%%%%%%
For magnetic fields of 20\,kG and more, there is an effect of
depolarization due to the blending of spectral lines. This
might explain why some stars with strong magnetic fields do not show
detectable signals of BBLP. 

The second assumption to be addressed is whether a dipolar field is
sufficient to approximate the magnetic field of CP stars. 

A natural way to describe a
(potential) magnetic field is to consider a multipolar expansion. From
a dipole, one can build up a quadrupolar field by considering two
opposite dipoles displaced by a vector which in general is not
parallel to the direction of the dipoles. The magnetic configuration
is specified by three directions (one for the dipole, and two for the
quadrupole) and by the strengths of the dipolar and of the quadrupolar
components. An advantage of this formalism is that many observable
quantities related to the magnetic field 
\textit{have an analytical representation} (Bagnulo et al.\ 1996).
The solid lines in Fig.~1 show the observable
quantities for a sample of magnetic configurations
described with a dipole plus a quadrupole. 
BBLP shows again to be the
most sensitive to the magnetic morphology. It should be noted that
the \textit{longitudinal magnetic field is not
sensitive to the quadrupolar component of the magnetic field}, unless
the quadrupole largely dominates the magnetic morphology. The
longitudinal magnetic field is described by a simple
sinusoidal curve, even when the magnetic field shows severe departures
from the dipolar geometry. Severe departures from the dipolar
morphology -- not detected via $\langle B_z\rangle$ 
measurements -- are probably responsible for the lack of consistency
between observed and predicted BBLP. 
The assumption that a dipolar field is sufficient to approximate the 
magnetic field of CP stars should be revisited.

In principle, a quick way to discover whether a star has a magnetic
field more complicated than the dipolar one is simply to plot 
$\langle B^2 + B_z^2\rangle^{1/2}$ vs.\ $\langle B_z\rangle$, which gives a
line if the magnetic field is dipolar and a (single) loop if the
field has a quadrupolar component. Figure~3 clearly
shows that the magnetic field of $\beta$~CrB exhibits severe
departures from the simple dipolar configuration. 
\newpage
%%%%%%%%%%%%%%%%%%%%%%%%%%%%%%%%%%%%%%%%%%%%%%%%%%%%%%%%%%%%%%%%%%%
\begin{figure}
\label{Fig_Beta}
%\begin{center}
%\scalebox{0.8}{
%\psfig{figure=T3f3.ps,width=11.3cm,clip=}}
%\includegraphics*[3.3cm,17.3cm][18.5cm,23.5cm]{Bagnulo_3.ps}}
\centerline{
\psfig{file=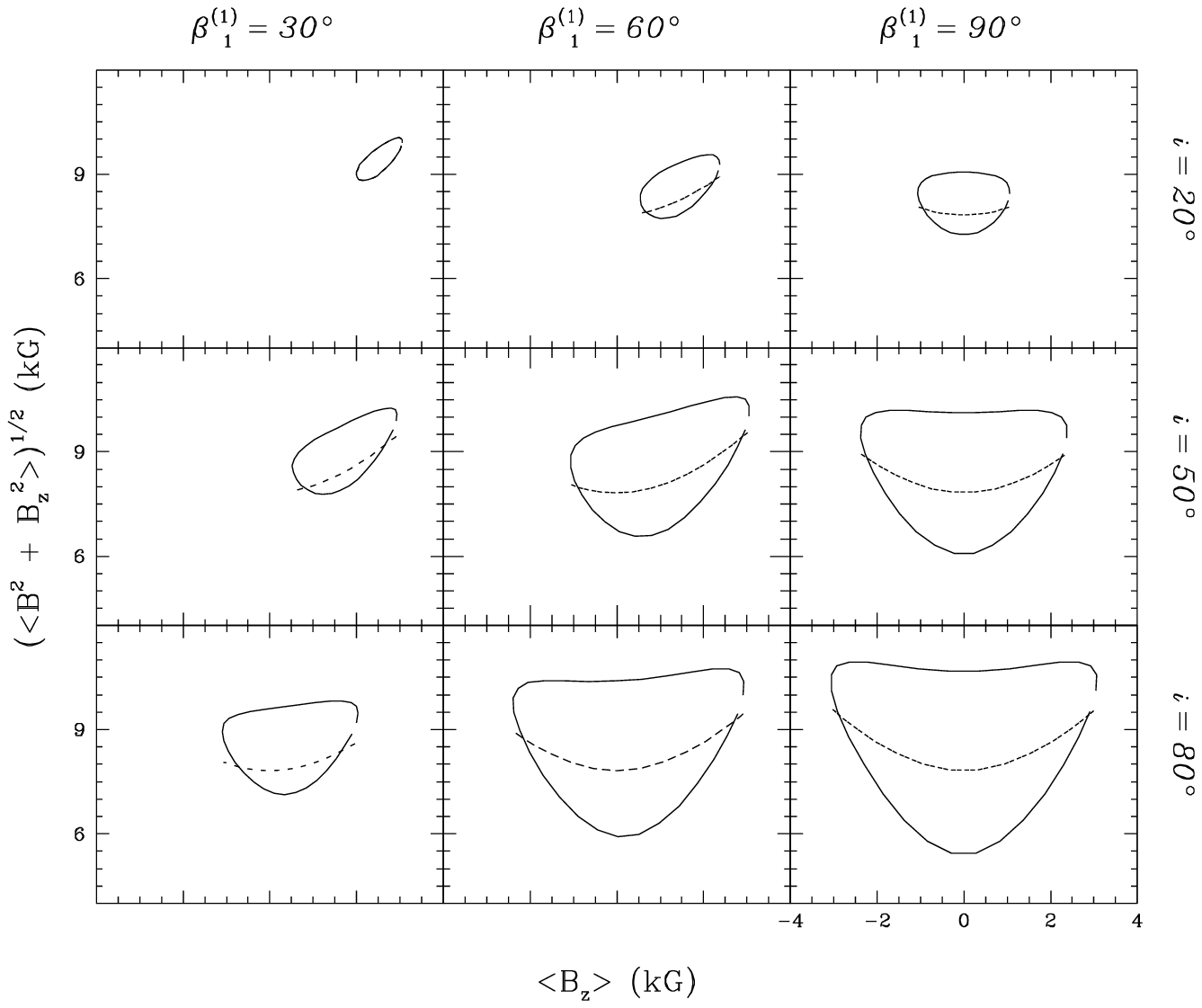,height=5.cm,clip=}
\hfill
\psfig{file=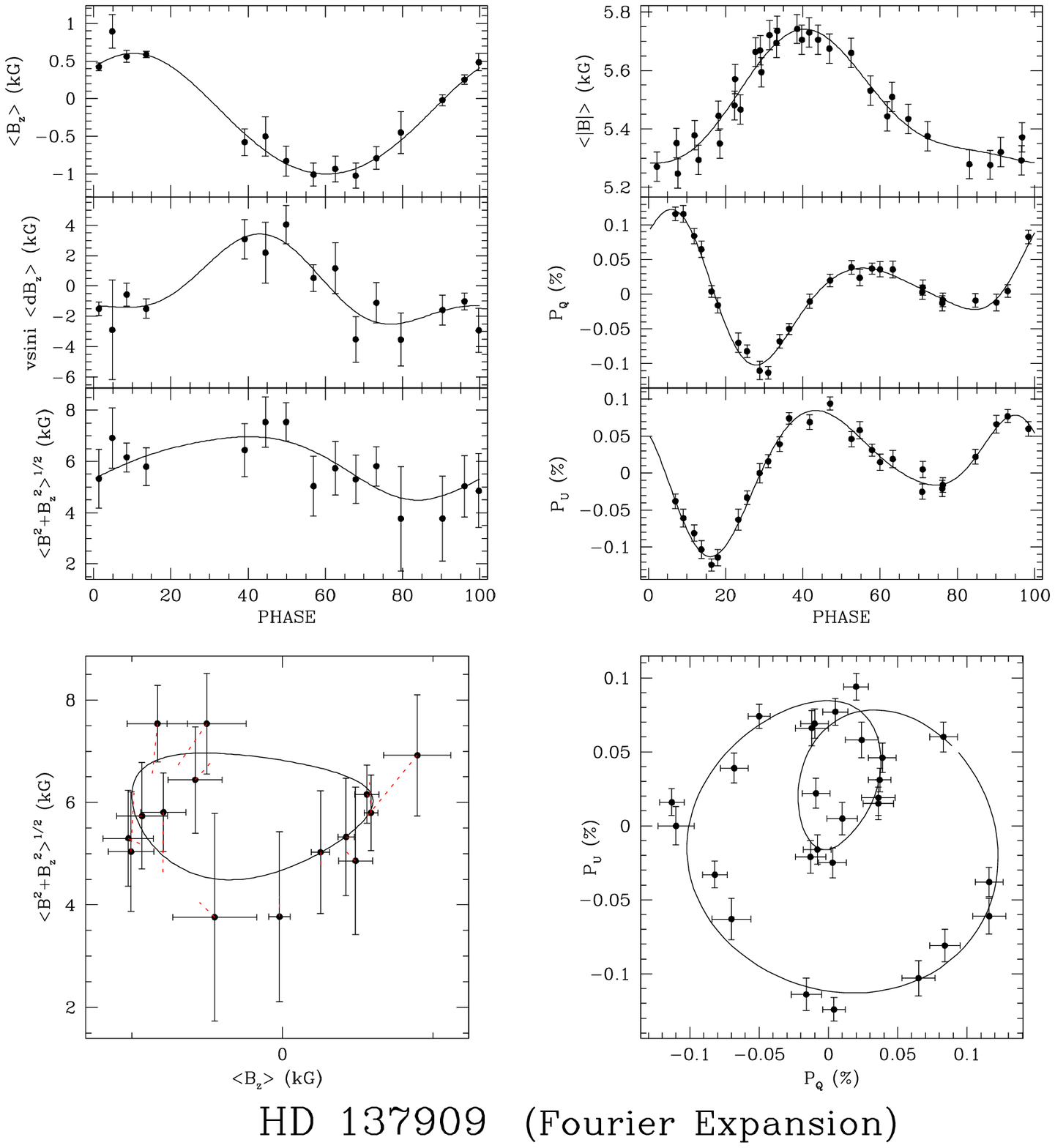,height=5.cm,clip=}
}
\caption{Clues to a quadrupolar magnetic field. Left: 
         $\langle B^2 + B_z^2 \rangle^{1/2}$ vs.\ $\langle B_z\rangle$ for a
         dipolar field (dotted lines) and dipole + quadrupole (solid
         lines). Departures from the dipolar morphology are responsible
	 for a loop in the $\langle B^2 + B^2_z \rangle^{1/2} -
	 \langle B_z \rangle$ plane.
	 Right: observations of $\beta$~CrB by
         Mathys (1994; 1995b) and Mathys \& Hubrig (1997). The solid line
         is a fit obtained with a Fourier expansion}
%\end{center}
\end{figure}
%%%%%%%%%%%%%%%%%%%%%%%%%%%%%%%%%%%%%%%%%%%%%%%%%%%%%%%%%%%%%%%%%%%

\acknowledgements
% Do not leave a blank line here! <---------------------->
This work has been supported by the Austrian {\em Fonds zur
F{\"o}rderung der Wissenschaftlichen Forschung}, project P12101-AST.

\end{document}